\documentclass[twocolumn,showpacs,superscriptaddress,prl]{revtex4-1}
\usepackage{graphicx}
\usepackage{xcolor}
\usepackage{amsmath,amsthm,amssymb}
\usepackage{braket}
\usepackage[colorlinks,citecolor=blue,linkcolor=red]{hyperref}
\usepackage{ulem}
\usepackage{soul}

\begin{document}
 
\title{An {\it ab initio} study of electron-hole pairs in a correlated van der Waals antiferromagnet: NiPS$_3$}

\author{Christopher Lane}
\email{laneca@lanl.gov}
\affiliation{Theoretical Division, Los Alamos National Laboratory, Los Alamos, New Mexico 87545, USA}
\affiliation{Center for Integrated Nanotechnologies, Los Alamos National Laboratory, Los Alamos, New Mexico 87545, USA}

\author{Jian-Xin Zhu}
\email{jxzhu@lanl.gov}
\affiliation{Theoretical Division, Los Alamos National Laboratory, Los Alamos, New Mexico 87545, USA}
\affiliation{Center for Integrated Nanotechnologies, Los Alamos National Laboratory, Los Alamos, New Mexico 87545, USA}

\date{\today} 
\begin{abstract}
The recently discovered two-dimensional (2D) van der Waals magnet NiPS$_3$ provides a new route to examine many-body quasiparticles under 2D confinement. Excitons are of particular interest due to their strong coupling to the magnetic ground state in this material. Here, by using a first-principles based approach, we find bright excitons between 1.4 eV to 1.7 eV, similar to the sharp coherent and band edge excitons experimentally observed. Our analysis shows that each exciton in NiPS$_3$ is composed of a combination of $d$-$d$ and charge-transfer character, where the relative ratio of each pairing configuration changes with exciton energy. Finally, the wave function of the electrons and holes is revealed to be spatially separated, with electrons and holes residing on different magnetic sublattices. This suggests a microscopic origin of the observed strong magneto-exciton coupling.
\end{abstract}


\maketitle 

\begin{figure}[t]
\includegraphics[width=0.98\columnwidth]{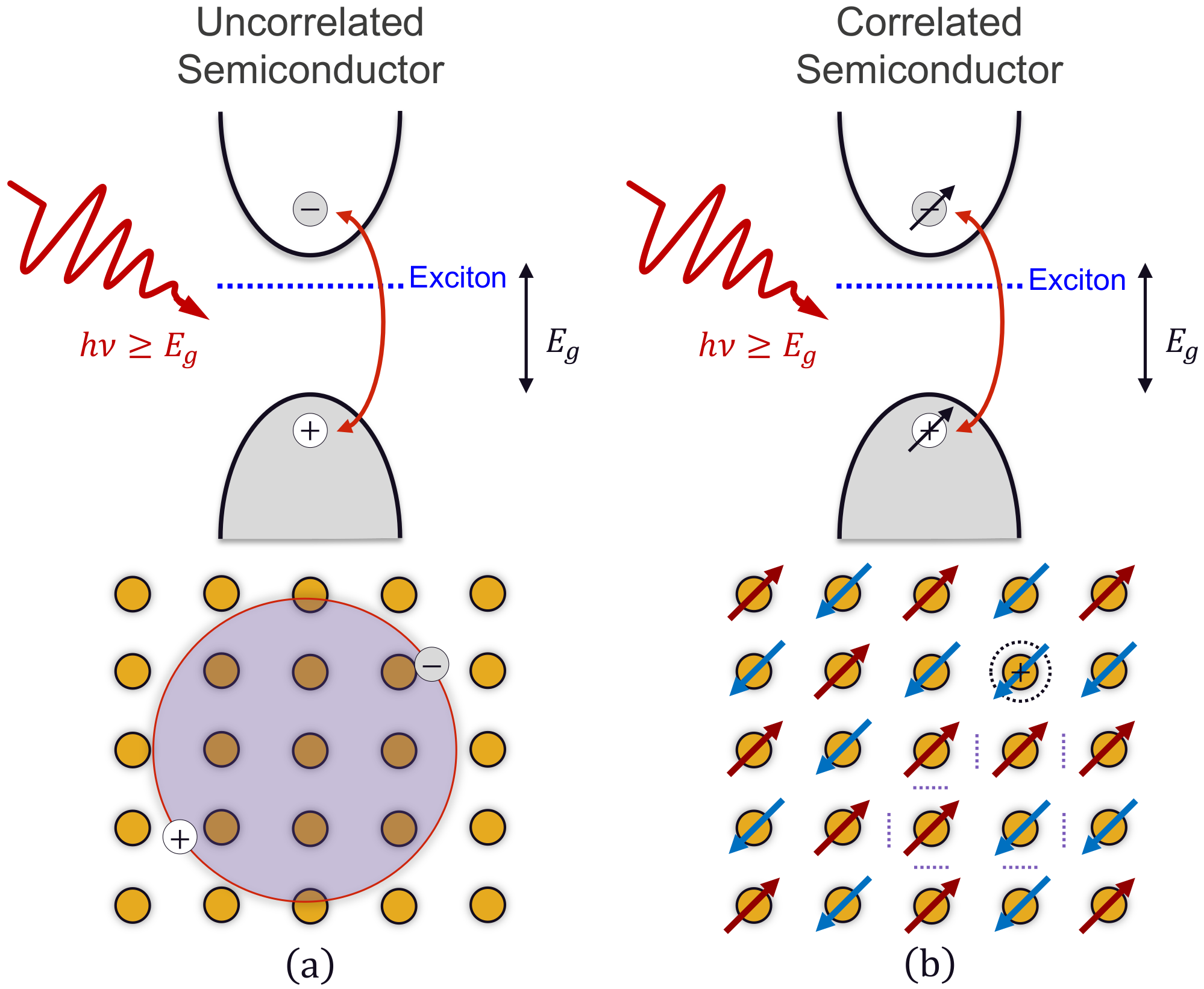}
\caption{(color online) (a) Excitons in band insulators move freely in a material. (b) In correlated materials excitons are intrinsically coupled to spin and orbital degrees of freedom allowing for the formation of more exotic composite quasiparticles. The illustration is equally valid the spin-down case.} 
\label{fig:SCHEMATIC}
\end{figure}

{\it Introduction.}---The emergence of excitons from a highly correlated ground state is an intriguing and a relatively unexplored problem in condensed matter physics.  Collective excitations of bound electron-hole pairs -- excitons -- are ubiquitous in band semiconductors \cite{aspnes1983dielectric,mueller2018exciton}, molecular crystals \cite{davydov1964theory,agranovich2009excitations}, and biological systems \cite{cao2020quantum}.  In simple band insulators [Fig.~\ref{fig:SCHEMATIC}(a)], electron-hole pairs are able to diffuse across the lattice and transport energy with no net electric charge.  In complex correlated materials, however, excitons are intrinsically interwoven into the complex emergent ground state properties, composed of several active physical interactions involving spin, charge, lattice, and orbital degrees of freedom [Fig.~\ref{fig:SCHEMATIC}(b)].  Therefore, these excitons not only offer a new playground for examining many-body phenomena beyond the physics of conventional insulators,  but also provide a direct window into the fundamental nature of the correlated insulating ground state. 

The recent discovery of two-dimensional (2D) van der Waals magnets presents a unique opportunity to study many-body excitons, owing to the strong Coulomb interactions and highly correlated ground states. In particular, NiPS$_3$ has gained attention for being a charge-transfer insulator with a zig-zag antiferromagntic ground state following the highly anisotropic XXZ  Heisenberg model \cite{wildes2022magnetic}. The ground state can be further tuned by thinning to yield strong magnetic fluctuations~\cite{kim2019suppression} and a possible Mott metal-insulator transition under pressure \cite{harms2022symmetry,kim2019mott}. These interesting properties resemble those of the high-$T_c$ cuprates, thus suggesting NiPS$_3$ may host unconventional superconductivity~\cite{gu2019ni,lane2020thickness}.

Beyond the ground state, NiPS$_3$ exhibits signatures of possible many-body excitons and strong correlation effects. Specifically, sharp exciton features in the optical spectra are suggested~\cite{kang2020coherent,ho2021band,hwangbo2021highly,belvin2021exciton} to be electron-hole bound states originating from $d$-$d$, charge-transfer, and localized many-body excitations, similar to Zhang–Rice singlets. Polarization and temperature dependent studies have revealed that these excitons strongly couple to various phonon modes, producing a diverse array of exciton-phonon bound states \cite{dirnberger2022spin,hwangbo2021highly,belvin2021exciton,ergeccen2022magnetically}, and are closely connected with the underlying stripe-like zig-zag antiferromagnetic order \cite{kang2020coherent,hwangbo2021highly}. Furthermore, by driving these dressed quasiparticles out of equilibrium, one can create a transient metallic state that preserves long-range antiferromagnetism~\cite{belvin2021exciton}, in addition to directly yielding the electron-phonon coupling strength~\cite{ergeccen2022magnetically} and controlling the magnetic anisotropy~\cite{afanasiev2021controlling}.

\begin{figure*}[ht]
\includegraphics[width=0.99\textwidth]{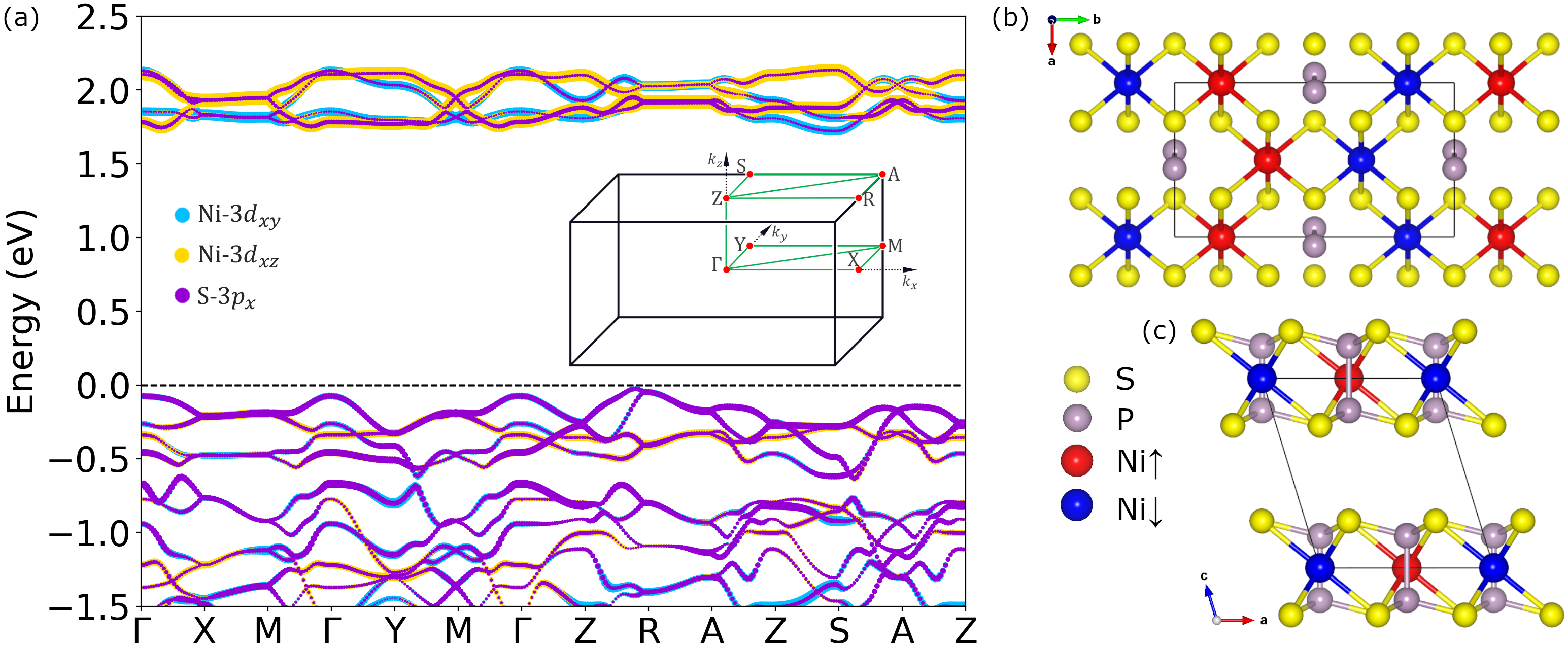}
\caption{(color online) (a) Electronic band structures (blue lines) along the high-symmetry lines in the Brillouin zone for NiPS$_3$ in the zig-zag AFM phase. Contribution of $d_{xy}$, $d_{xz}$, and $p_{x}$  orbitals are highlighted with light blue, gold, and violet dots, respectively. The sizes of the dots are proportional to the fractional weights of orbital species. A schematic diagram of the Brillouin zones with the path followed in presenting the band structures is shown on the right.   (b) and (c) Theoretically predicted zig-zag AFM state of NiPS$_3$ in the  monoclinic crystal structure. The related zig-zag AFM structure is highlighted by coloring the nickel atoms in NiPS$_3$ red (blue) for spin-up (down). The black lines mark the unit cell.} 
\label{fig:BANDSanndSTRUCTURE}
\end{figure*}

Despite these compelling excited state properties, theoretical studies have been limited. To date, first-principles based calculations have focused on the electronic and magentic ground state \cite{kim2019mott,olsen2021magnetic,lane2020thickness,birowska2021large}, magnetic anisotropy \cite{olsen2021magnetic,kim2021magnetic}, and the optical properties within the independent particle approximation \cite{lane2020thickness}. Therefore, a detailed study of the various excitonic states and their character is still missing. 

In this Letter, we present the exciton states of NiPS$_3$ in the zig-zag AFM phase for zero and finite momentum transfer. We capture the key experimentally observed excitons spanning 1.4 eV to 1.7 eV. Specifically, we predict an extremely bright exciton at 1.434 eV similar to the coherent many-body exciton observed at 1.47 eV. The exciton dispersion exhibits a moderate bandwidth for excitations in all three directions, indicating a significant interlayer coupling in the bulk compound. By projecting the electron-hole coupling matrix onto the atomic-sites using the full manifold of atomic orbitals, we find the excitons to be composed of a combination of pairing configurations, including $d$-$d$ and charge-transfer. Furthermore, the relative ratio between pairing arrangements changes from predominately $d$-$d$ to charge-transfer as the energy of the exciton increases, inline with experimental reports. Finally, by examining the exciton wave functions, we find the electrons and holes to reside on opposite magentic sublattices. The existence of inter-magnetic sublattice excitons suggests the presence of a strong magneto-exciton coupling in accord with experiments.

{\it Computational Details.}---First-principles calculations were carried out by using the pseudopotential projector-augmented wave (PAW) method \cite{Kresse1999} implemented in the Vienna ab initio simulation package (VASP) \cite{Kresse1996,Kresse1993} with an energy cutoff of $400$ eV for the plane-wave basis set. The GW PAW potentials released with VASP.5.4 were used. Exchange-correlation effects were treated using the SCAN meta-GGA scheme \cite{Sun2015}. A 7 $\times$ 3 $\times$ 7 $\Gamma$-centered k-point mesh was used to sample the Brillouin zone. Due to very weak spin-orbit coupling exhibited by the relatively light nickel atoms, spin-orbit coupling was neglected. We used the experimentally obtained atomic positions and lattice parameters for the bulk C2/m (space group number 12) structure throughout this work \cite{taylor1973preparation}. A total energy tolerance of $10^{-8}$ eV was used to determine the self-consistent charge density. The response functions and exciton eigenvalue calculations followed the procedure of Ref.~\onlinecite{lane2020landscape} using 96 frequency points and 600 virtual orbitals with an energy cutoff equal to half of the plane-wave cutoff. To obtain excitons at zero and finite center-of-mass momentum, $\mathbf{Q}$, the Bethe-Salpetter equation (BSE) was solved beyond the Tamm-Dancoff approximation. Since SCAN is constructed within the generalized Kohn-Sham scheme~\cite{seidl1996generalized} and our recent SCAN-based study of NiPS$_3$ obtains a band gap in good accord with experimental observations~\cite{lane2020thickness}, we avoid the GW quasiparticle corrections and directly use the generalized Kohn-Sham band energies as the eigenvalues of the electrons and holes in the BSE Hamiltonian, where only 8 conduction and 16 valence bands. 

\begin{figure*}[ht]
\includegraphics[width=0.99\textwidth]{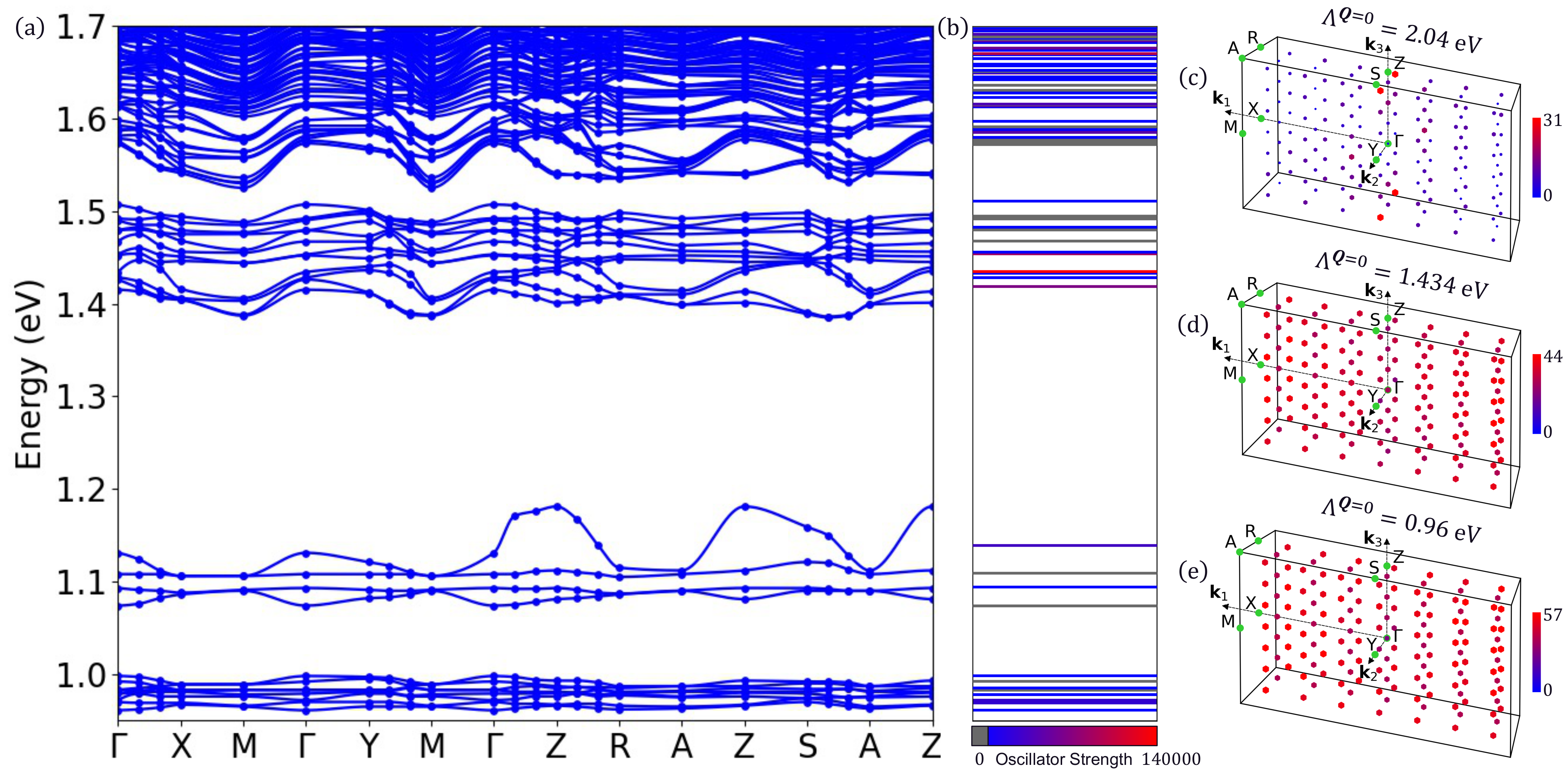}
\caption{(color online) (a) Exciton band dispersion (blue dots and lines) along the high-symmetry lines in the Brillouin zone of momentum transfer for NiPS$_3$ in the zig-zag AFM phase. (b) The exciton state energy levels for $\mathbf{Q}=0$, where the color goes as the oscillator strength associated with a given exciton. The energy scale is the same as (a). (c)-(e) The contribution of each momenta $\mathbf{k}$ in first Brillouin zone for three excitons of different energies $\Lambda^{\mathbf{Q}}$ for $\mathbf{Q}=0$.}
\label{fig:EXCBANDS}
\end{figure*}

{\it Ground State Properties.}---Figure~\ref{fig:BANDSanndSTRUCTURE}(a) shows the electronic band dispersion and (b)-(c) the crystal structure of bulk NiPS$_3$ in the zig-zag AFM phase.  The site-resolved atomic projections for Ni-$3d_{xy}$ (teal dots),  Ni-$3d_{xz}$ (gold dots),  and S-$p_x$ (purple dots) are overlaid  where the size of the dots are proportional to the fractional orbital weights. Since each nickel atom has a $2^+$ ionic state, due to the highly covalent P$_2$S$_6$ anion complex of $4^-$ charge, a zig-zag AFM state [Fig.~\ref{fig:BANDSanndSTRUCTURE}(b)] is stabilized across the nickel atoms.  As a result of strong electron-electron interactions,  the nickel-dominated states bookend the full bandwidth of Ni-S hybridized levels. Thus producing a band gap of 1.719 eV, with a predominantly Ni conduction band (82\%) and valance bands composed of mostly S-$p$ orbital character (66\%).  This arrangement of states is typical of a charge-transfer insulator \cite{zaanen1985band}, similar to the cuprates and the Ruddlesden-Popper perovskite nickelates. For more details and a thorough study of the ground-state magnetic and electronic structure of NiPS$_3$ employing the SCAN functional please refer to Ref.~\onlinecite{lane2020thickness}.

To understand the electron-hole pairs in NiPS$_3$,  we focus on the low energy electronic structure near the valence and conduction band edges. The band gap is indirect, with the lowest energetic transition occurring at $\approx(\frac{\pi}{4},0,\pi)$ and $(0,\pi,\pi)$, for the valence and conduction bands,  respectively. Since the system breaks four-fold rotational symmetry in the $ab$ plane,  the X(R) and Y(S) directions in the Brillouin zone are inequivalent. This results in the states near Y(S) to be shifted below those at X(R) by $\sim$ $0.25$ eV.  Turning to the atomic-site resolved band projections, the Ni-$3d_{xy}$ and Ni-$3d_{xz}$ projected states in the conduction band appear to be localized on two separate bands. For all k-points along the high-symmetry path except for X$-$M the $d_{xz}$ states are lower in energy by $\approx 2-10$ meV. Sulfur-$p_x$  orbital weight plays a minimal role in these bands throughout the zone. The valence band character is predominantly sulfur $p_x$, with little variation along the path in momentum space.  A non-negligible admixture of $d_{xz}$/$d_{xy}$ is also present due to the strong hybridization between nickel and sulfur atoms.

\begin{figure*}[ht!]
\includegraphics[width=0.99\textwidth]{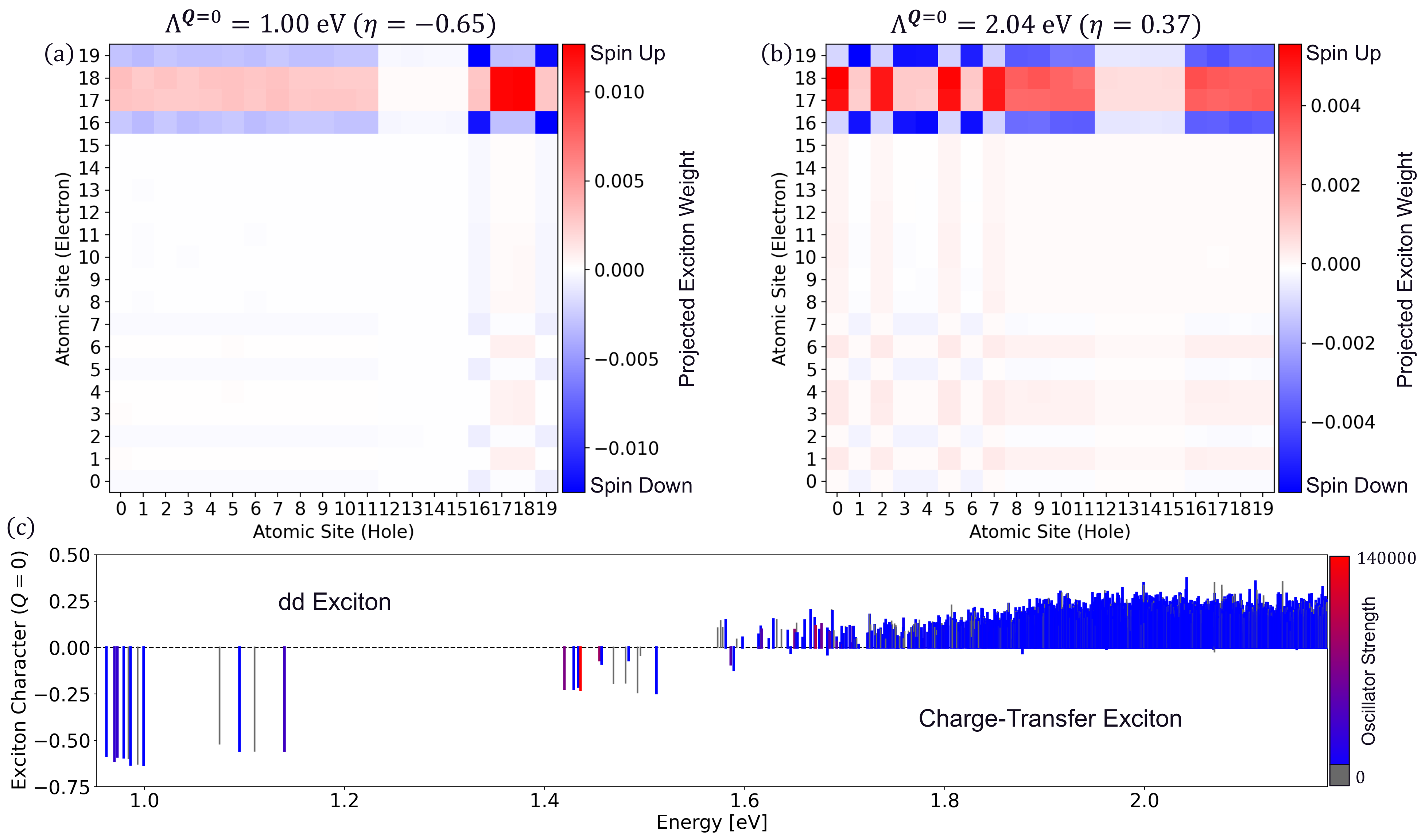}
\caption{(color online) Heat map of the atomic-site projected electron-hole coupling amplitude, $|C^{S}_{\tau, \tau^{\prime}}|^2$, for a (a) low and (b) high energy exciton in NiPS$_3$. The positions of the various atoms in the unit cell are given in the Supplemental Material. (c) Character mixing parameter ($\eta^{S\mathbf{Q}=0}$) as a function of exciton energy. The color goes as the oscillator strength associated with a given exciton.} 
\label{fig:PROJEXC}
\end{figure*}

{\it Excitonic Properties.}---Figure~\ref{fig:EXCBANDS}(a) shows the energy dispersion of the first 80 exciton states along the high symmetry directions in the first Brillouin zone of bulk NiPS$_3$. Two clusters of low-energy exciton bands,  comprised of 4 and 6 excitons,  respectively,  are found separated from the remaining continuum of excitonic states by $\sim 400$ meV. The exciton states appear to be an admixture of relatively localized narrow bands at low energies and highly mobile dispersive excitons for higher energies. All excitons seem to be mobile despite the zig-zag AFM background. Additionally, the appreciable $q_z$ dispersion suggests electron-hole pairs can readily hop between van der Waals layers owing to the significant interlayer coupling. In particular, states near 1.1 eV display an enhanced dispersion along the $z$-axis and in the Z-plane compared to other momentum values near $\Gamma$. Interestingly, the low lying exciton bands follow a parabolic line shape similar to those found in single-layer cuprates \cite{lane2020landscape}.

The oscillator strength, or {\it brightness}, reveals a few optically $(\mathbf{Q}=0)$ relevant electron-hole pairs [Fig.~\ref{fig:EXCBANDS}(b)]. We find a very bright (135000 oscillator strength) exciton at 1.434 eV similar to the strong coherent many-body exciton observed at 1.47 eV  \cite{kang2020coherent,hwangbo2021highly,ho2021band,belvin2021exciton}. Other moderately bright states are found at 1.455 eV, 1.650 eV and 1.671 eV, in accord with the band-edge, spin-orbit coupled, and $d$-$d$ excitons reported in Refs.~\onlinecite{kang2020coherent,ho2021band,hwangbo2021highly,belvin2021exciton}. A large number of the remaining exciton states are optically active, however with significantly reduced oscillator strength.

Figure~\ref{fig:EXCBANDS}(c)-(e) show the contribution of each momenta $\mathbf{k}$ in first Brillouin zone for three excitons of increasing energy $\Lambda^{\mathbf{Q}}$ (0.96 eV, 1.434 eV, and 2.04 eV) with $\mathbf{Q}=0$ to give insight into the energy dependence of exciton localization. The size and color of each dot goes as  $\sum_{cv\sigma}  |Z^{S\sigma\mathbf{Q}}_{cv\mathbf{k}}|^2$, where $Z^{S\sigma\mathbf{Q}}_{cv\mathbf{k}}$ is the electron-hole amplitude \cite{sander2015beyond,lane2020landscape}, $S$ indexes the exciton state, $\mathbf{k}$ ($\mathbf{Q}$) is the crystal momentum (center-of-mass momentum), and the contribution of each band, valence $v$ and conduction $c$, and electron spin $\sigma$ is integrated out. In NiPS$_3$, the momentum distribution associated with a given exciton appears to depend sensitively on its energy. For the lowest energy exciton (0.96 eV), the weight achieves a maximum at the Brillouin zone edge,  with a $\sim$50\% dip in amplitude as $\mathbf{k}$ approaches the three planes that pass through $\Gamma$. The moderately uniform weight distribution in momentum space stems from the relatively flat conduction bands and narrow valance states. In turn,  the lowest energy exciton is reasonably localized in real space and moderately mobile,  as reflected by its dispersion. 

\begin{figure}[t]
\includegraphics[width=0.90\columnwidth]{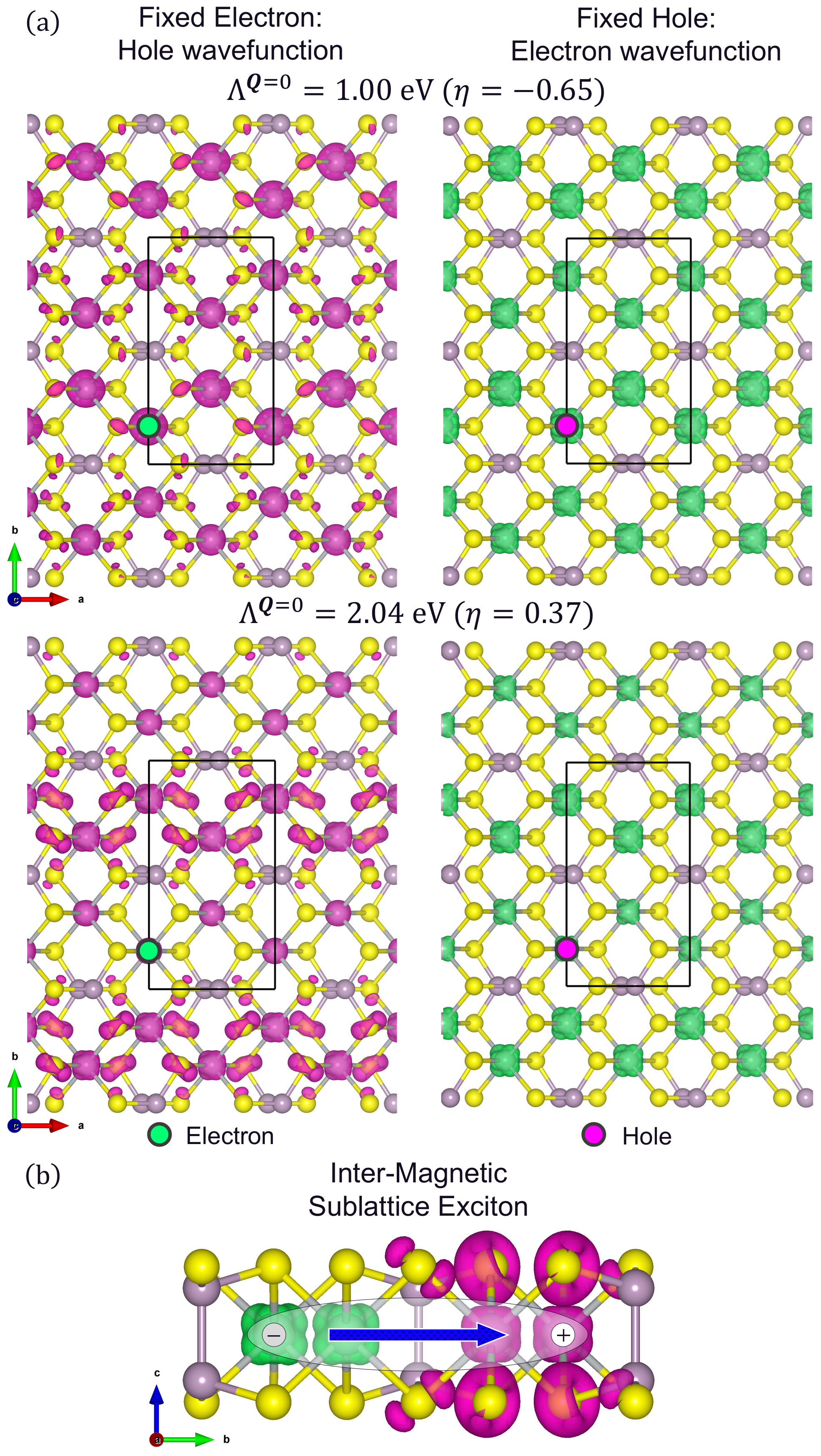}
\caption{(color online) (a) The wave function of a predominately $d$-$d$  and charge-transer exciton in NiPS$_3$ for a fixed electron (light green dot) and hole (pink dot). The nickel, sulfur and phosphorus atoms are represented by silver, yellow, and violet spheres, respectively. The black lines mark the unit cell. (b) A cross section of an inter-magnetic sublattice exciton, with the charge centers and induced electric dipole moment indicated.} 
\label{fig:WF}
\end{figure}

The exceptionally bright exciton at 1.434 eV is similarly delocalized in momentum space as the lowest energy exciton,  with only a slight increase in the overall uniformity of the weight distribution throughout the zone [Fig.~\ref{fig:EXCBANDS} (d)].  Additionally, this exciton shows an enhanced dispersion, suggesting this electron-hole pair can more easily move about the sample. Finally, as the exciton energies increase, we find a stark change in localization behavior.  Specifically, for an exciton just above the band edge (2.04 eV) the contribution of each momenta $\mathbf{k}$ in first Brillouin zone becomes highly localized, suggesting the electrons and holes have become very loosely bound and are more delocalized within the solid.  

Important to classifying the various exciton states and disentangling the various contributing degrees of freedom, is to ask where the electrons and holes reside within the material system. To address this question, we follow Ref.~\onlinecite{lane2020landscape} and project $Z^{S\sigma}_{cv\mathbf{k}}$ onto each atomic site including the full manifold of atomic orbitals corresponding to the particular atomic species, yielding,
\begin{align}
C^{S\sigma\mathbf{Q}}_{\tau lm , \tau^{\prime} l^{\prime}m^{\prime}}= \sum_{\mathbf{k}cv\sigma} Z^{S\sigma\mathbf{Q}}_{cv\mathbf{k}}   P^{\mathbf{k+\mathbf{Q}}c\sigma}_{\tau lm}  P^{*~\mathbf{k}v\sigma}_{\tau^{\prime} l^{\prime}m^{\prime} } ,
\end{align}
where $\tau$ indexes the site,  $lm$ specify the real spherical harmonic $Y_{lm}$,  and $P^{\mathbf{k}n\sigma}_{\tau lm} $ is the transformation between bases. Due to the multiplicity of active orbitals at the valence (conduction) band edge, we further simplify the discussion by integrating out the orbital degrees of freedom,  $C^{S\sigma\mathbf{Q}}_{\tau, \tau^{\prime}}$.  We note that $C^{S\sigma\mathbf{Q}}_{\tau, \tau^{\prime}}$ exhibits minimal overlap between spin-up and spin-down atomic site matrix elements.

Figure~\ref{fig:PROJEXC} shows a heat map of $\sum_{\sigma}\left(-1\right)^{\sigma}|C^{S\sigma\mathbf{Q}=0}_{\tau, \tau^{\prime}}|^2$ for the (a) 1.00 eV and (b) 2.04 eV excitons in NiPS$_3$,  where the horizontal and vertical axes are the atomic sites of the holes and electrons, respectively.  For the 1.00 eV exciton, a clear asymmetry about the diagonal is present, indicating a difference in the most likely atomic position of the electron and hole.  For example, the electron has a higher probability of sitting on the nickel atomic sites as compared to the hole which displays an additional weaker weight across the sulfur atoms.  Overall, the nickel-nickel sector exhibits the largest amplitude for exciton formation, where the highest probability  matrix elements corresponds to pairing within the same magnetic sublattice, suggesting the existence of $d$-$d$ type excitons in NiPS$_3$.  Interestingly,  there is non-negligible probability for an exciton to form between the two nickel atoms of opposing magnetization, giving rise to possible Mott-Hubbard excitons.  Additional non-zero weight is found in the Ni-S sector signaling the presents of charge-transfer electron-hole pairs.  

For the high-energy exciton [Fig.~\ref{fig:PROJEXC}(b)], we find a different arrangement of atomic character. Specifically, the electron-hole pairs exhibit a similar clear asymmetry about the diagonal. However, the nickel-nickel sector is no longer the dominant pairing pathway. In this case, the Ni-S sector provides the largest pairing amplitude and predicts excitons with electrons and holes residing on different magnetic sublattices. This sublattice selectivity is directly inline with reports of strong magneto-exciton coupling~\cite{ergeccen2022magnetically,kang2020coherent,ho2021band,hwangbo2021highly,belvin2021exciton}. These charge-transfer excitons dominate the spectrum, with the $d$-$d$ configurations playing a reduced role. Finally, due to the strong covalent bonding between sulfur and phosphorous, a finite phosphorous atomic character also appears.

By comparing low and high-energy excitons, there is a clear change in the prevailing pairing arrangement. To quantify this observation, we define the {\it character mixing} parameter $\eta^{S\mathbf{Q}}$ for a given exciton state $S$ of momentum $\mathbf{Q}$ as 
\begin{align}
\eta^{S\mathbf{Q}} = \frac{\sum_{\sigma} |C^{S\sigma\mathbf{Q}}_{Ni,Ni}|^2  -  \sum_{\sigma} |C^{S\sigma\mathbf{Q}}_{Ni,S}|^2 }{\sum_{\sigma} |C^{S\sigma\mathbf{Q}}_{Ni,Ni}|^2  + \sum_{\sigma} |C^{S\sigma\mathbf{Q}}_{Ni,S}|^2 }
\end{align}
where $|C^{S\sigma\mathbf{Q}}_{Ni,Ni}|^2 ~ \left( |C^{S\sigma\mathbf{Q}}_{Ni,S}|^2 \right)$ is the total amplitude of the  nickel-nickel (nickel-sulfur) sector.  Figure~\ref{fig:PROJEXC}(c) presents the character mixing of the first 3000 exciton states at $\mathbf{Q}=0$. The color is proportional to the oscillator strength (brightness) of each exciton. As can be seen,  the character of the various excitonic states transitions from predominantly $d$-$d$ type for low-energies to charge-transfer type for high-energies with the transition occurring just before the band edge at $\approx 1.6$ eV. Interestingly, the brightest excitons in NiPS$_3$ emerge in a region of strong mixing $(|\eta|\lesssim0.20)$ between $d$-$d$ and charge-transfer pairing types. Similar changes in character has been observed in optical spectroscopy and resonant inelastic x-ray spectroscopy (RIXS) \cite{kang2020coherent,belvin2021exciton}. 

To visualize the anisotropy of the exciton wave function brought about by separating electrons and holes on opposite magnetic sublattices, we plot the exciton wave function for both low- and high-energy excitons. To represent the six-coordinate function, 
\begin{align}
\Psi^{S\mathbf{Q}}(\mathbf{r}_e,\mathbf{r}_h)=\sum_{\mathbf{k}cv\sigma} Z^{S\sigma\mathbf{Q}}_{cv\mathbf{k}}   \phi_{c\mathbf{k+Q}}(\mathbf{r}_e)\phi^*_{v\mathbf{k}}(\mathbf{r}_h)  \;,
\end{align}
we fix the hole (electron) position and plot the resulting electron (hole) density, e.g., $|\Psi^{S}(\mathbf{r}_e,\mathbf{r}_h=\mathbf{R})|^2$. Figure~\ref{fig:WF} (a) shows the hole and electron density of the exciton wave function for 1.0 eV (top panel) and 2.04 eV (bottom panel) excitons. Fixing the electron on a nickel atomic site (light green dot), the 1.0 eV exciton has an associated hole wave function that is largely localized on the nickel atoms within the same magnetic sublattice, with lesser density spread out over the opposite magneticly polarized sites. In contrast, the electron wave function for a fixed hole (pink dot) on a nickel site is tightly localized to the nickel atomic centers on both magnetic lattices. As the exciton energy increases, the hole density (fixed electron) increases on the sulfur atoms of the opposite magnetic sublattice, while the weight on the nickel atoms diminishes. Concomitantly, the electron density (fixed hole) gradually shifts weight from intra- to inter-magnetic sublattice. These trends follow the observation of Fig.~\ref{fig:PROJEXC}(c) showing a transition from $d$-$d$ to charge-transfer electron-hole pairs with energy. Moreover, the highly anisotropic spatial extent of the exciton wave function, spanning both magnetic sublattices, further corroborates reports of strong AFM coupling and mixed character~\cite{kang2020coherent,hwangbo2021highly,belvin2021exciton,ergeccen2022magnetically}.

The appearance of inter-magnetic sublattice excitons is quite reminiscent of interlayer electron-hole pairs. In transition metal dichalcogenide van der Waals heterostructures, a type-II band alignment facilitates interlayer charge transfer with electrons accumulating in the layer with a lower conduction band minimum and holes collecting in the other layer with the higher valence band maximum~\cite{jiang2021interlayer}. Analogously, the alternating magnetic polarization of the zig-zag AFM order ensures the valance and conduction bands of equivalent spins are isolated on differing magnetic sublattices, thereby producing electron-hole pairs with spatially separated electron (hole) wave functions, as shown in Fig.~\ref{fig:WF}(b). Consequently, a local electric dipole is induced at the interface between magnetic sublattices whose direction is perpendicular to the zigzag-AFM order~[Fig.~\ref{fig:WF}(b)]. Furthermore, these excess hole (electron) carriers transferred to the magnetic sites by photoexcitation effectively dope the local zigzag AFM order. This gives way to local metallic puddles and persistent spin-wave excitations, though of reduced strength, in accord with recent pump-probe observations~\cite{belvin2021exciton} and other doped strongly correlated materials~\cite{horigane2016spin,le2013dispersive,le2011intense}. The overlap of the electron and hole wave functions also plays a crucial role in determining exciton lifetime~\cite{knox1963theory,jiang2021interlayer}. Interlayer excitons, for example, exhibit lifetimes on the order of nanoseconds~\cite{miller2017long} or larger~\cite{jiang2018microsecond} due to the sizable interlayer separation in van der Waals heterostrcures. For NiPS$_3$, the closely interwoven magnetic sublattices yields a wave function overlap greater than interlayer excitons. This suggests the lifetime of the sharp 1.47 eV exciton maybe increased from picoseconds~\cite{wang2021spin}, if the distance between magnetic sublattices were to be increased, e.g. via strain. This would allow for enhanced electrical control and manipulation of optical and transport properties, in addition to the magneto-exciton coupling already observed~\cite{kang2020coherent,hwangbo2021highly}.

{\it Concluding Remarks.}---Our study has demonstrated that NiPS$_3$ supports a number of electron-hole pairing pathways, including $d$-$d$ and charge-transfer, in agreement with experimental reports. An exceptionally bright exciton is found to emerge at 1.434 eV with strong mixing between $d$-$d$ and charge-transfer character. Interestingly, the spatial separation of electron and hole wave functions, suggests inter-magnetic sublattice excitons may provide an alternative novel information carrier for hybrid electrical-optical microelectronic systems, due to its ability to being coupled to both optical and magnetic perturbations. Finally, our study gives insight into the nature of excitons in strongly correlated materials, and stimulates further experimental study of excitons in AFM systems.


\begin{acknowledgments}
This work was carried out under the auspices of the U.S. Department of Energy (DOE) National Nuclear Security Administration under Contract No. 89233218CNA000001. It was supported by the LANL LDRD Program, the Quantum Science Center, a U.S. DOE Office of Science National Quantum Information Science Research Center, and in part by the Center for Integrated Nanotechnologies, a DOE BES user facility, in partnership with the LANL Institutional Computing Program for computational resources. Additional computations were performed at the National Energy Research Scientific Computing Center (NERSC), a U.S. Department of Energy Office of Science User Facility located at Lawrence Berkeley National Laboratory, operated under Contract No. DE-AC02-05CH11231 using NERSC award ERCAP0020494. 
\end{acknowledgments} 

\bibliography{Refs}

\begin{thebibliography}{38}%
\makeatletter
\providecommand \@ifxundefined [1]{%
 \@ifx{#1\undefined}
}%
\providecommand \@ifnum [1]{%
 \ifnum #1\expandafter \@firstoftwo
 \else \expandafter \@secondoftwo
 \fi
}%
\providecommand \@ifx [1]{%
 \ifx #1\expandafter \@firstoftwo
 \else \expandafter \@secondoftwo
 \fi
}%
\providecommand \natexlab [1]{#1}%
\providecommand \enquote  [1]{``#1''}%
\providecommand \bibnamefont  [1]{#1}%
\providecommand \bibfnamefont [1]{#1}%
\providecommand \citenamefont [1]{#1}%
\providecommand \href@noop [0]{\@secondoftwo}%
\providecommand \href [0]{\begingroup \@sanitize@url \@href}%
\providecommand \@href[1]{\@@startlink{#1}\@@href}%
\providecommand \@@href[1]{\endgroup#1\@@endlink}%
\providecommand \@sanitize@url [0]{\catcode `\\12\catcode `\$12\catcode
  `\&12\catcode `\#12\catcode `\^12\catcode `\_12\catcode `\%12\relax}%
\providecommand \@@startlink[1]{}%
\providecommand \@@endlink[0]{}%
\providecommand \url  [0]{\begingroup\@sanitize@url \@url }%
\providecommand \@url [1]{\endgroup\@href {#1}{\urlprefix }}%
\providecommand \urlprefix  [0]{URL }%
\providecommand \Eprint [0]{\href }%
\providecommand \doibase [0]{http://dx.doi.org/}%
\providecommand \selectlanguage [0]{\@gobble}%
\providecommand \bibinfo  [0]{\@secondoftwo}%
\providecommand \bibfield  [0]{\@secondoftwo}%
\providecommand \translation [1]{[#1]}%
\providecommand \BibitemOpen [0]{}%
\providecommand \bibitemStop [0]{}%
\providecommand \bibitemNoStop [0]{.\EOS\space}%
\providecommand \EOS [0]{\spacefactor3000\relax}%
\providecommand \BibitemShut  [1]{\csname bibitem#1\endcsname}%
\let\auto@bib@innerbib\@empty
\bibitem [{\citenamefont {Aspnes}\ and\ \citenamefont
  {Studna}(1983)}]{aspnes1983dielectric}%
  \BibitemOpen
  \bibfield  {author} {\bibinfo {author} {\bibfnamefont {D.~E.}\ \bibnamefont
  {Aspnes}}\ and\ \bibinfo {author} {\bibfnamefont {A.}~\bibnamefont
  {Studna}},\ }\href@noop {} {\bibfield  {journal} {\bibinfo  {journal}
  {Physical Review B}\ }\textbf {\bibinfo {volume} {27}},\ \bibinfo {pages}
  {985} (\bibinfo {year} {1983})}\BibitemShut {NoStop}%
\bibitem [{\citenamefont {Mueller}\ and\ \citenamefont
  {Malic}(2018)}]{mueller2018exciton}%
  \BibitemOpen
  \bibfield  {author} {\bibinfo {author} {\bibfnamefont {T.}~\bibnamefont
  {Mueller}}\ and\ \bibinfo {author} {\bibfnamefont {E.}~\bibnamefont
  {Malic}},\ }\href@noop {} {\bibfield  {journal} {\bibinfo  {journal} {npj 2D
  Materials and Applications}\ }\textbf {\bibinfo {volume} {2}},\ \bibinfo
  {pages} {1} (\bibinfo {year} {2018})}\BibitemShut {NoStop}%
\bibitem [{\citenamefont {Davydov}(1964)}]{davydov1964theory}%
  \BibitemOpen
  \bibfield  {author} {\bibinfo {author} {\bibfnamefont {A.~S.}\ \bibnamefont
  {Davydov}},\ }\href@noop {} {\bibfield  {journal} {\bibinfo  {journal}
  {Soviet Physics Uspekhi}\ }\textbf {\bibinfo {volume} {7}},\ \bibinfo {pages}
  {145} (\bibinfo {year} {1964})}\BibitemShut {NoStop}%
\bibitem [{\citenamefont {Agranovich}(2009)}]{agranovich2009excitations}%
  \BibitemOpen
  \bibfield  {author} {\bibinfo {author} {\bibfnamefont {V.~M.}\ \bibnamefont
  {Agranovich}},\ }\href@noop {} {\emph {\bibinfo {title} {Excitations in
  organic solids}}},\ Vol.\ \bibinfo {volume} {142}\ (\bibinfo  {publisher}
  {OUP Oxford},\ \bibinfo {year} {2009})\BibitemShut {NoStop}%
\bibitem [{\citenamefont {Cao}\ \emph {et~al.}(2020)\citenamefont {Cao},
  \citenamefont {Cogdell}, \citenamefont {Coker}, \citenamefont {Duan},
  \citenamefont {Hauer}, \citenamefont {Kleinekath{\"o}fer}, \citenamefont
  {Jansen}, \citenamefont {Man{\v{c}}al}, \citenamefont {Miller}, \citenamefont
  {Ogilvie} \emph {et~al.}}]{cao2020quantum}%
  \BibitemOpen
  \bibfield  {author} {\bibinfo {author} {\bibfnamefont {J.}~\bibnamefont
  {Cao}}, \bibinfo {author} {\bibfnamefont {R.~J.}\ \bibnamefont {Cogdell}},
  \bibinfo {author} {\bibfnamefont {D.~F.}\ \bibnamefont {Coker}}, \bibinfo
  {author} {\bibfnamefont {H.-G.}\ \bibnamefont {Duan}}, \bibinfo {author}
  {\bibfnamefont {J.}~\bibnamefont {Hauer}}, \bibinfo {author} {\bibfnamefont
  {U.}~\bibnamefont {Kleinekath{\"o}fer}}, \bibinfo {author} {\bibfnamefont
  {T.~L.}\ \bibnamefont {Jansen}}, \bibinfo {author} {\bibfnamefont
  {T.}~\bibnamefont {Man{\v{c}}al}}, \bibinfo {author} {\bibfnamefont {R.~D.}\
  \bibnamefont {Miller}}, \bibinfo {author} {\bibfnamefont {J.~P.}\
  \bibnamefont {Ogilvie}},  \emph {et~al.},\ }\href@noop {} {\bibfield
  {journal} {\bibinfo  {journal} {Science Advances}\ }\textbf {\bibinfo
  {volume} {6}},\ \bibinfo {pages} {eaaz4888} (\bibinfo {year}
  {2020})}\BibitemShut {NoStop}%
\bibitem [{\citenamefont {Wildes}\ \emph {et~al.}(2022)\citenamefont {Wildes},
  \citenamefont {Stewart}, \citenamefont {Le}, \citenamefont {Ewings},
  \citenamefont {Ruke}, \citenamefont {Deng},\ and\ \citenamefont
  {Anand}}]{wildes2022magnetic}%
  \BibitemOpen
  \bibfield  {author} {\bibinfo {author} {\bibfnamefont {A.}~\bibnamefont
  {Wildes}}, \bibinfo {author} {\bibfnamefont {J.}~\bibnamefont {Stewart}},
  \bibinfo {author} {\bibfnamefont {M.}~\bibnamefont {Le}}, \bibinfo {author}
  {\bibfnamefont {R.}~\bibnamefont {Ewings}}, \bibinfo {author} {\bibfnamefont
  {K.}~\bibnamefont {Ruke}}, \bibinfo {author} {\bibfnamefont {G.}~\bibnamefont
  {Deng}}, \ and\ \bibinfo {author} {\bibfnamefont {K.}~\bibnamefont {Anand}},\
  }\href@noop {} {\bibfield  {journal} {\bibinfo  {journal} {arXiv preprint
  arXiv:2207.07448}\ } (\bibinfo {year} {2022})}\BibitemShut {NoStop}%
\bibitem [{\citenamefont {Kim}\ \emph {et~al.}(2019{\natexlab{a}})\citenamefont
  {Kim}, \citenamefont {Lim}, \citenamefont {Lee}, \citenamefont {Lee},
  \citenamefont {Kim}, \citenamefont {Park}, \citenamefont {Jeon},
  \citenamefont {Park}, \citenamefont {Park},\ and\ \citenamefont
  {Cheong}}]{kim2019suppression}%
  \BibitemOpen
  \bibfield  {author} {\bibinfo {author} {\bibfnamefont {K.}~\bibnamefont
  {Kim}}, \bibinfo {author} {\bibfnamefont {S.~Y.}\ \bibnamefont {Lim}},
  \bibinfo {author} {\bibfnamefont {J.-U.}\ \bibnamefont {Lee}}, \bibinfo
  {author} {\bibfnamefont {S.}~\bibnamefont {Lee}}, \bibinfo {author}
  {\bibfnamefont {T.~Y.}\ \bibnamefont {Kim}}, \bibinfo {author} {\bibfnamefont
  {K.}~\bibnamefont {Park}}, \bibinfo {author} {\bibfnamefont {G.~S.}\
  \bibnamefont {Jeon}}, \bibinfo {author} {\bibfnamefont {C.-H.}\ \bibnamefont
  {Park}}, \bibinfo {author} {\bibfnamefont {J.-G.}\ \bibnamefont {Park}}, \
  and\ \bibinfo {author} {\bibfnamefont {H.}~\bibnamefont {Cheong}},\
  }\href@noop {} {\bibfield  {journal} {\bibinfo  {journal} {Nature
  Communications}\ }\textbf {\bibinfo {volume} {10}},\ \bibinfo {pages} {345}
  (\bibinfo {year} {2019}{\natexlab{a}})}\BibitemShut {NoStop}%
\bibitem [{\citenamefont {Harms}\ \emph {et~al.}(2022)\citenamefont {Harms},
  \citenamefont {Matsuoka}, \citenamefont {Samanta}, \citenamefont {Clune},
  \citenamefont {Smith}, \citenamefont {Haglund}, \citenamefont {Feng},
  \citenamefont {Cao}, \citenamefont {Smith}, \citenamefont {Mandrus} \emph
  {et~al.}}]{harms2022symmetry}%
  \BibitemOpen
  \bibfield  {author} {\bibinfo {author} {\bibfnamefont {N.~C.}\ \bibnamefont
  {Harms}}, \bibinfo {author} {\bibfnamefont {T.}~\bibnamefont {Matsuoka}},
  \bibinfo {author} {\bibfnamefont {S.}~\bibnamefont {Samanta}}, \bibinfo
  {author} {\bibfnamefont {A.~J.}\ \bibnamefont {Clune}}, \bibinfo {author}
  {\bibfnamefont {K.~A.}\ \bibnamefont {Smith}}, \bibinfo {author}
  {\bibfnamefont {A.~V.}\ \bibnamefont {Haglund}}, \bibinfo {author}
  {\bibfnamefont {E.}~\bibnamefont {Feng}}, \bibinfo {author} {\bibfnamefont
  {H.}~\bibnamefont {Cao}}, \bibinfo {author} {\bibfnamefont {J.~S.}\
  \bibnamefont {Smith}}, \bibinfo {author} {\bibfnamefont {D.~G.}\ \bibnamefont
  {Mandrus}},  \emph {et~al.},\ }\href@noop {} {\bibfield  {journal} {\bibinfo
  {journal} {npj 2D Materials and Applications}\ }\textbf {\bibinfo {volume}
  {6}},\ \bibinfo {pages} {40} (\bibinfo {year} {2022})}\BibitemShut {NoStop}%
\bibitem [{\citenamefont {Kim}\ \emph {et~al.}(2019{\natexlab{b}})\citenamefont
  {Kim}, \citenamefont {Haule},\ and\ \citenamefont
  {Vanderbilt}}]{kim2019mott}%
  \BibitemOpen
  \bibfield  {author} {\bibinfo {author} {\bibfnamefont {H.-S.}\ \bibnamefont
  {Kim}}, \bibinfo {author} {\bibfnamefont {K.}~\bibnamefont {Haule}}, \ and\
  \bibinfo {author} {\bibfnamefont {D.}~\bibnamefont {Vanderbilt}},\
  }\href@noop {} {\bibfield  {journal} {\bibinfo  {journal} {Physical Review
  Letters}\ }\textbf {\bibinfo {volume} {123}},\ \bibinfo {pages} {236401}
  (\bibinfo {year} {2019}{\natexlab{b}})}\BibitemShut {NoStop}%
\bibitem [{\citenamefont {Gu}\ \emph {et~al.}(2019)\citenamefont {Gu},
  \citenamefont {Zhang}, \citenamefont {Le}, \citenamefont {Li}, \citenamefont
  {Xiang},\ and\ \citenamefont {Hu}}]{gu2019ni}%
  \BibitemOpen
  \bibfield  {author} {\bibinfo {author} {\bibfnamefont {Y.}~\bibnamefont
  {Gu}}, \bibinfo {author} {\bibfnamefont {Q.}~\bibnamefont {Zhang}}, \bibinfo
  {author} {\bibfnamefont {C.}~\bibnamefont {Le}}, \bibinfo {author}
  {\bibfnamefont {Y.}~\bibnamefont {Li}}, \bibinfo {author} {\bibfnamefont
  {T.}~\bibnamefont {Xiang}}, \ and\ \bibinfo {author} {\bibfnamefont
  {J.}~\bibnamefont {Hu}},\ }\href@noop {} {\bibfield  {journal} {\bibinfo
  {journal} {Physical Review B}\ }\textbf {\bibinfo {volume} {100}},\ \bibinfo
  {pages} {165405} (\bibinfo {year} {2019})}\BibitemShut {NoStop}%
\bibitem [{\citenamefont {Lane}\ and\ \citenamefont
  {Zhu}(2020{\natexlab{a}})}]{lane2020thickness}%
  \BibitemOpen
  \bibfield  {author} {\bibinfo {author} {\bibfnamefont {C.}~\bibnamefont
  {Lane}}\ and\ \bibinfo {author} {\bibfnamefont {J.-X.}\ \bibnamefont {Zhu}},\
  }\href@noop {} {\bibfield  {journal} {\bibinfo  {journal} {Physical Review
  B}\ }\textbf {\bibinfo {volume} {102}},\ \bibinfo {pages} {075124} (\bibinfo
  {year} {2020}{\natexlab{a}})}\BibitemShut {NoStop}%
\bibitem [{\citenamefont {Kang}\ \emph {et~al.}(2020)\citenamefont {Kang},
  \citenamefont {Kim}, \citenamefont {Kim}, \citenamefont {Kim}, \citenamefont
  {Sim}, \citenamefont {Lee}, \citenamefont {Lee}, \citenamefont {Park},
  \citenamefont {Yun}, \citenamefont {Kim} \emph {et~al.}}]{kang2020coherent}%
  \BibitemOpen
  \bibfield  {author} {\bibinfo {author} {\bibfnamefont {S.}~\bibnamefont
  {Kang}}, \bibinfo {author} {\bibfnamefont {K.}~\bibnamefont {Kim}}, \bibinfo
  {author} {\bibfnamefont {B.~H.}\ \bibnamefont {Kim}}, \bibinfo {author}
  {\bibfnamefont {J.}~\bibnamefont {Kim}}, \bibinfo {author} {\bibfnamefont
  {K.~I.}\ \bibnamefont {Sim}}, \bibinfo {author} {\bibfnamefont {J.-U.}\
  \bibnamefont {Lee}}, \bibinfo {author} {\bibfnamefont {S.}~\bibnamefont
  {Lee}}, \bibinfo {author} {\bibfnamefont {K.}~\bibnamefont {Park}}, \bibinfo
  {author} {\bibfnamefont {S.}~\bibnamefont {Yun}}, \bibinfo {author}
  {\bibfnamefont {T.}~\bibnamefont {Kim}},  \emph {et~al.},\ }\href@noop {}
  {\bibfield  {journal} {\bibinfo  {journal} {Nature}\ }\textbf {\bibinfo
  {volume} {583}},\ \bibinfo {pages} {785} (\bibinfo {year}
  {2020})}\BibitemShut {NoStop}%
\bibitem [{\citenamefont {Ho}\ \emph {et~al.}(2021)\citenamefont {Ho},
  \citenamefont {Hsu},\ and\ \citenamefont {Muhimmah}}]{ho2021band}%
  \BibitemOpen
  \bibfield  {author} {\bibinfo {author} {\bibfnamefont {C.-H.}\ \bibnamefont
  {Ho}}, \bibinfo {author} {\bibfnamefont {T.-Y.}\ \bibnamefont {Hsu}}, \ and\
  \bibinfo {author} {\bibfnamefont {L.~C.}\ \bibnamefont {Muhimmah}},\
  }\href@noop {} {\bibfield  {journal} {\bibinfo  {journal} {npj 2D Materials
  and Applications}\ }\textbf {\bibinfo {volume} {5}},\ \bibinfo {pages} {1}
  (\bibinfo {year} {2021})}\BibitemShut {NoStop}%
\bibitem [{\citenamefont {Hwangbo}\ \emph {et~al.}(2021)\citenamefont
  {Hwangbo}, \citenamefont {Zhang}, \citenamefont {Jiang}, \citenamefont
  {Wang}, \citenamefont {Fonseca}, \citenamefont {Wang}, \citenamefont
  {Diederich}, \citenamefont {Gamelin}, \citenamefont {Xiao}, \citenamefont
  {Chu} \emph {et~al.}}]{hwangbo2021highly}%
  \BibitemOpen
  \bibfield  {author} {\bibinfo {author} {\bibfnamefont {K.}~\bibnamefont
  {Hwangbo}}, \bibinfo {author} {\bibfnamefont {Q.}~\bibnamefont {Zhang}},
  \bibinfo {author} {\bibfnamefont {Q.}~\bibnamefont {Jiang}}, \bibinfo
  {author} {\bibfnamefont {Y.}~\bibnamefont {Wang}}, \bibinfo {author}
  {\bibfnamefont {J.}~\bibnamefont {Fonseca}}, \bibinfo {author} {\bibfnamefont
  {C.}~\bibnamefont {Wang}}, \bibinfo {author} {\bibfnamefont {G.~M.}\
  \bibnamefont {Diederich}}, \bibinfo {author} {\bibfnamefont {D.~R.}\
  \bibnamefont {Gamelin}}, \bibinfo {author} {\bibfnamefont {D.}~\bibnamefont
  {Xiao}}, \bibinfo {author} {\bibfnamefont {J.-H.}\ \bibnamefont {Chu}},
  \emph {et~al.},\ }\href@noop {} {\bibfield  {journal} {\bibinfo  {journal}
  {Nature Nanotechnology}\ }\textbf {\bibinfo {volume} {16}},\ \bibinfo {pages}
  {655} (\bibinfo {year} {2021})}\BibitemShut {NoStop}%
\bibitem [{\citenamefont {Belvin}\ \emph {et~al.}(2021)\citenamefont {Belvin},
  \citenamefont {Baldini}, \citenamefont {Ozel}, \citenamefont {Mao},
  \citenamefont {Po}, \citenamefont {Allington}, \citenamefont {Son},
  \citenamefont {Kim}, \citenamefont {Kim}, \citenamefont {Hwang} \emph
  {et~al.}}]{belvin2021exciton}%
  \BibitemOpen
  \bibfield  {author} {\bibinfo {author} {\bibfnamefont {C.~A.}\ \bibnamefont
  {Belvin}}, \bibinfo {author} {\bibfnamefont {E.}~\bibnamefont {Baldini}},
  \bibinfo {author} {\bibfnamefont {I.~O.}\ \bibnamefont {Ozel}}, \bibinfo
  {author} {\bibfnamefont {D.}~\bibnamefont {Mao}}, \bibinfo {author}
  {\bibfnamefont {H.~C.}\ \bibnamefont {Po}}, \bibinfo {author} {\bibfnamefont
  {C.~J.}\ \bibnamefont {Allington}}, \bibinfo {author} {\bibfnamefont
  {S.}~\bibnamefont {Son}}, \bibinfo {author} {\bibfnamefont {B.~H.}\
  \bibnamefont {Kim}}, \bibinfo {author} {\bibfnamefont {J.}~\bibnamefont
  {Kim}}, \bibinfo {author} {\bibfnamefont {I.}~\bibnamefont {Hwang}},  \emph
  {et~al.},\ }\href@noop {} {\bibfield  {journal} {\bibinfo  {journal} {Nature
  Communications}\ }\textbf {\bibinfo {volume} {12}},\ \bibinfo {pages} {4837}
  (\bibinfo {year} {2021})}\BibitemShut {NoStop}%
\bibitem [{\citenamefont {Dirnberger}\ \emph {et~al.}(2022)\citenamefont
  {Dirnberger}, \citenamefont {Bushati}, \citenamefont {Datta}, \citenamefont
  {Kumar}, \citenamefont {MacDonald}, \citenamefont {Baldini},\ and\
  \citenamefont {Menon}}]{dirnberger2022spin}%
  \BibitemOpen
  \bibfield  {author} {\bibinfo {author} {\bibfnamefont {F.}~\bibnamefont
  {Dirnberger}}, \bibinfo {author} {\bibfnamefont {R.}~\bibnamefont {Bushati}},
  \bibinfo {author} {\bibfnamefont {B.}~\bibnamefont {Datta}}, \bibinfo
  {author} {\bibfnamefont {A.}~\bibnamefont {Kumar}}, \bibinfo {author}
  {\bibfnamefont {A.~H.}\ \bibnamefont {MacDonald}}, \bibinfo {author}
  {\bibfnamefont {E.}~\bibnamefont {Baldini}}, \ and\ \bibinfo {author}
  {\bibfnamefont {V.~M.}\ \bibnamefont {Menon}},\ }\href@noop {} {\bibfield
  {journal} {\bibinfo  {journal} {Nature Nanotechnology}\ ,\ \bibinfo {pages}
  {1}} (\bibinfo {year} {2022})}\BibitemShut {NoStop}%
\bibitem [{\citenamefont {Erge{\c{c}}en}\ \emph {et~al.}(2022)\citenamefont
  {Erge{\c{c}}en}, \citenamefont {Ilyas}, \citenamefont {Mao}, \citenamefont
  {Po}, \citenamefont {Yilmaz}, \citenamefont {Kim}, \citenamefont {Park},
  \citenamefont {Senthil},\ and\ \citenamefont
  {Gedik}}]{ergeccen2022magnetically}%
  \BibitemOpen
  \bibfield  {author} {\bibinfo {author} {\bibfnamefont {E.}~\bibnamefont
  {Erge{\c{c}}en}}, \bibinfo {author} {\bibfnamefont {B.}~\bibnamefont
  {Ilyas}}, \bibinfo {author} {\bibfnamefont {D.}~\bibnamefont {Mao}}, \bibinfo
  {author} {\bibfnamefont {H.~C.}\ \bibnamefont {Po}}, \bibinfo {author}
  {\bibfnamefont {M.~B.}\ \bibnamefont {Yilmaz}}, \bibinfo {author}
  {\bibfnamefont {J.}~\bibnamefont {Kim}}, \bibinfo {author} {\bibfnamefont
  {J.-G.}\ \bibnamefont {Park}}, \bibinfo {author} {\bibfnamefont
  {T.}~\bibnamefont {Senthil}}, \ and\ \bibinfo {author} {\bibfnamefont
  {N.}~\bibnamefont {Gedik}},\ }\href@noop {} {\bibfield  {journal} {\bibinfo
  {journal} {Nature Communications}\ }\textbf {\bibinfo {volume} {13}},\
  \bibinfo {pages} {98} (\bibinfo {year} {2022})}\BibitemShut {NoStop}%
\bibitem [{\citenamefont {Afanasiev}\ \emph {et~al.}(2021)\citenamefont
  {Afanasiev}, \citenamefont {Hortensius}, \citenamefont {Matthiesen},
  \citenamefont {Ma{\~n}as-Valero}, \citenamefont {{\v{S}}i{\v{s}}kins},
  \citenamefont {Lee}, \citenamefont {Lesne}, \citenamefont {van Der~Zant},
  \citenamefont {Steeneken}, \citenamefont {Ivanov} \emph
  {et~al.}}]{afanasiev2021controlling}%
  \BibitemOpen
  \bibfield  {author} {\bibinfo {author} {\bibfnamefont {D.}~\bibnamefont
  {Afanasiev}}, \bibinfo {author} {\bibfnamefont {J.~R.}\ \bibnamefont
  {Hortensius}}, \bibinfo {author} {\bibfnamefont {M.}~\bibnamefont
  {Matthiesen}}, \bibinfo {author} {\bibfnamefont {S.}~\bibnamefont
  {Ma{\~n}as-Valero}}, \bibinfo {author} {\bibfnamefont {M.}~\bibnamefont
  {{\v{S}}i{\v{s}}kins}}, \bibinfo {author} {\bibfnamefont {M.}~\bibnamefont
  {Lee}}, \bibinfo {author} {\bibfnamefont {E.}~\bibnamefont {Lesne}}, \bibinfo
  {author} {\bibfnamefont {H.~S.}\ \bibnamefont {van Der~Zant}}, \bibinfo
  {author} {\bibfnamefont {P.~G.}\ \bibnamefont {Steeneken}}, \bibinfo {author}
  {\bibfnamefont {B.~A.}\ \bibnamefont {Ivanov}},  \emph {et~al.},\ }\href@noop
  {} {\bibfield  {journal} {\bibinfo  {journal} {Science Advances}\ }\textbf
  {\bibinfo {volume} {7}},\ \bibinfo {pages} {eabf3096} (\bibinfo {year}
  {2021})}\BibitemShut {NoStop}%
\bibitem [{\citenamefont {Olsen}(2021)}]{olsen2021magnetic}%
  \BibitemOpen
  \bibfield  {author} {\bibinfo {author} {\bibfnamefont {T.}~\bibnamefont
  {Olsen}},\ }\href@noop {} {\bibfield  {journal} {\bibinfo  {journal} {Journal
  of Physics D: Applied Physics}\ }\textbf {\bibinfo {volume} {54}},\ \bibinfo
  {pages} {314001} (\bibinfo {year} {2021})}\BibitemShut {NoStop}%
\bibitem [{\citenamefont {Birowska}\ \emph {et~al.}(2021)\citenamefont
  {Birowska}, \citenamefont {Junior}, \citenamefont {Fabian},\ and\
  \citenamefont {Kunstmann}}]{birowska2021large}%
  \BibitemOpen
  \bibfield  {author} {\bibinfo {author} {\bibfnamefont {M.}~\bibnamefont
  {Birowska}}, \bibinfo {author} {\bibfnamefont {P.~E.~F.}\ \bibnamefont
  {Junior}}, \bibinfo {author} {\bibfnamefont {J.}~\bibnamefont {Fabian}}, \
  and\ \bibinfo {author} {\bibfnamefont {J.}~\bibnamefont {Kunstmann}},\
  }\href@noop {} {\bibfield  {journal} {\bibinfo  {journal} {Physical Review
  B}\ }\textbf {\bibinfo {volume} {103}},\ \bibinfo {pages} {L121108} (\bibinfo
  {year} {2021})}\BibitemShut {NoStop}%
\bibitem [{\citenamefont {Kim}\ and\ \citenamefont
  {Park}(2021)}]{kim2021magnetic}%
  \BibitemOpen
  \bibfield  {author} {\bibinfo {author} {\bibfnamefont {T.~Y.}\ \bibnamefont
  {Kim}}\ and\ \bibinfo {author} {\bibfnamefont {C.-H.}\ \bibnamefont {Park}},\
  }\href@noop {} {\bibfield  {journal} {\bibinfo  {journal} {Nano Letters}\
  }\textbf {\bibinfo {volume} {21}},\ \bibinfo {pages} {10114} (\bibinfo {year}
  {2021})}\BibitemShut {NoStop}%
\bibitem [{\citenamefont {Kresse}\ and\ \citenamefont
  {Joubert}(1999)}]{Kresse1999}%
  \BibitemOpen
  \bibfield  {author} {\bibinfo {author} {\bibfnamefont {G.}~\bibnamefont
  {Kresse}}\ and\ \bibinfo {author} {\bibfnamefont {D.}~\bibnamefont
  {Joubert}},\ }\href@noop {} {\bibfield  {journal} {\bibinfo  {journal}
  {Physical Review B}\ }\textbf {\bibinfo {volume} {59}},\ \bibinfo {pages}
  {1758} (\bibinfo {year} {1999})}\BibitemShut {NoStop}%
\bibitem [{\citenamefont {Kresse}\ and\ \citenamefont
  {Furthm{\"{u}}ller}(1996)}]{Kresse1996}%
  \BibitemOpen
  \bibfield  {author} {\bibinfo {author} {\bibfnamefont {G.}~\bibnamefont
  {Kresse}}\ and\ \bibinfo {author} {\bibfnamefont {J.}~\bibnamefont
  {Furthm{\"{u}}ller}},\ }\href@noop {} {\bibfield  {journal} {\bibinfo
  {journal} {Physical Review B}\ }\textbf {\bibinfo {volume} {54}},\ \bibinfo
  {pages} {11169} (\bibinfo {year} {1996})}\BibitemShut {NoStop}%
\bibitem [{\citenamefont {Kresse}\ and\ \citenamefont
  {Hafner}(1993)}]{Kresse1993}%
  \BibitemOpen
  \bibfield  {author} {\bibinfo {author} {\bibfnamefont {G.}~\bibnamefont
  {Kresse}}\ and\ \bibinfo {author} {\bibfnamefont {J.}~\bibnamefont
  {Hafner}},\ }\href@noop {} {\bibfield  {journal} {\bibinfo  {journal}
  {Physical Review B}\ }\textbf {\bibinfo {volume} {48}},\ \bibinfo {pages}
  {13115} (\bibinfo {year} {1993})}\BibitemShut {NoStop}%
\bibitem [{\citenamefont {Sun}\ \emph {et~al.}(2015)\citenamefont {Sun},
  \citenamefont {Ruzsinszky},\ and\ \citenamefont {Perdew}}]{Sun2015}%
  \BibitemOpen
  \bibfield  {author} {\bibinfo {author} {\bibfnamefont {J.}~\bibnamefont
  {Sun}}, \bibinfo {author} {\bibfnamefont {A.}~\bibnamefont {Ruzsinszky}}, \
  and\ \bibinfo {author} {\bibfnamefont {J.}~\bibnamefont {Perdew}},\
  }\href@noop {} {\bibfield  {journal} {\bibinfo  {journal} {Physical Review
  Letters}\ }\textbf {\bibinfo {volume} {115}},\ \bibinfo {pages} {036402}
  (\bibinfo {year} {2015})}\BibitemShut {NoStop}%
\bibitem [{\citenamefont {Taylor}\ \emph {et~al.}(1973)\citenamefont {Taylor},
  \citenamefont {Steger},\ and\ \citenamefont {Wold}}]{taylor1973preparation}%
  \BibitemOpen
  \bibfield  {author} {\bibinfo {author} {\bibfnamefont {B.~E.}\ \bibnamefont
  {Taylor}}, \bibinfo {author} {\bibfnamefont {J.}~\bibnamefont {Steger}}, \
  and\ \bibinfo {author} {\bibfnamefont {A.}~\bibnamefont {Wold}},\ }\href@noop
  {} {\bibfield  {journal} {\bibinfo  {journal} {Journal of Solid State
  Chemistry}\ }\textbf {\bibinfo {volume} {7}},\ \bibinfo {pages} {461}
  (\bibinfo {year} {1973})}\BibitemShut {NoStop}%
\bibitem [{\citenamefont {Lane}\ and\ \citenamefont
  {Zhu}(2020{\natexlab{b}})}]{lane2020landscape}%
  \BibitemOpen
  \bibfield  {author} {\bibinfo {author} {\bibfnamefont {C.}~\bibnamefont
  {Lane}}\ and\ \bibinfo {author} {\bibfnamefont {J.-X.}\ \bibnamefont {Zhu}},\
  }\href@noop {} {\bibfield  {journal} {\bibinfo  {journal} {Physical Review
  B}\ }\textbf {\bibinfo {volume} {101}},\ \bibinfo {pages} {155135} (\bibinfo
  {year} {2020}{\natexlab{b}})}\BibitemShut {NoStop}%
\bibitem [{\citenamefont {Seidl}\ \emph {et~al.}(1996)\citenamefont {Seidl},
  \citenamefont {G{\"o}rling}, \citenamefont {Vogl}, \citenamefont {Majewski},\
  and\ \citenamefont {Levy}}]{seidl1996generalized}%
  \BibitemOpen
  \bibfield  {author} {\bibinfo {author} {\bibfnamefont {A.}~\bibnamefont
  {Seidl}}, \bibinfo {author} {\bibfnamefont {A.}~\bibnamefont {G{\"o}rling}},
  \bibinfo {author} {\bibfnamefont {P.}~\bibnamefont {Vogl}}, \bibinfo {author}
  {\bibfnamefont {J.}~\bibnamefont {Majewski}}, \ and\ \bibinfo {author}
  {\bibfnamefont {M.}~\bibnamefont {Levy}},\ }\href@noop {} {\bibfield
  {journal} {\bibinfo  {journal} {Physical Review B}\ }\textbf {\bibinfo
  {volume} {53}},\ \bibinfo {pages} {3764} (\bibinfo {year}
  {1996})}\BibitemShut {NoStop}%
\bibitem [{\citenamefont {Zaanen}\ \emph {et~al.}(1985)\citenamefont {Zaanen},
  \citenamefont {Sawatzky},\ and\ \citenamefont {Allen}}]{zaanen1985band}%
  \BibitemOpen
  \bibfield  {author} {\bibinfo {author} {\bibfnamefont {J.}~\bibnamefont
  {Zaanen}}, \bibinfo {author} {\bibfnamefont {G.}~\bibnamefont {Sawatzky}}, \
  and\ \bibinfo {author} {\bibfnamefont {J.}~\bibnamefont {Allen}},\
  }\href@noop {} {\bibfield  {journal} {\bibinfo  {journal} {Physical Review
  Letters}\ }\textbf {\bibinfo {volume} {55}},\ \bibinfo {pages} {418}
  (\bibinfo {year} {1985})}\BibitemShut {NoStop}%
\bibitem [{\citenamefont {Sander}\ \emph {et~al.}(2015)\citenamefont {Sander},
  \citenamefont {Maggio},\ and\ \citenamefont {Kresse}}]{sander2015beyond}%
  \BibitemOpen
  \bibfield  {author} {\bibinfo {author} {\bibfnamefont {T.}~\bibnamefont
  {Sander}}, \bibinfo {author} {\bibfnamefont {E.}~\bibnamefont {Maggio}}, \
  and\ \bibinfo {author} {\bibfnamefont {G.}~\bibnamefont {Kresse}},\
  }\href@noop {} {\bibfield  {journal} {\bibinfo  {journal} {Physical Review
  B}\ }\textbf {\bibinfo {volume} {92}},\ \bibinfo {pages} {045209} (\bibinfo
  {year} {2015})}\BibitemShut {NoStop}%
\bibitem [{\citenamefont {Jiang}\ \emph {et~al.}(2021)\citenamefont {Jiang},
  \citenamefont {Chen}, \citenamefont {Zheng}, \citenamefont {Zheng},\ and\
  \citenamefont {Pan}}]{jiang2021interlayer}%
  \BibitemOpen
  \bibfield  {author} {\bibinfo {author} {\bibfnamefont {Y.}~\bibnamefont
  {Jiang}}, \bibinfo {author} {\bibfnamefont {S.}~\bibnamefont {Chen}},
  \bibinfo {author} {\bibfnamefont {W.}~\bibnamefont {Zheng}}, \bibinfo
  {author} {\bibfnamefont {B.}~\bibnamefont {Zheng}}, \ and\ \bibinfo {author}
  {\bibfnamefont {A.}~\bibnamefont {Pan}},\ }\href@noop {} {\bibfield
  {journal} {\bibinfo  {journal} {Light: Science \& Applications}\ }\textbf
  {\bibinfo {volume} {10}},\ \bibinfo {pages} {1} (\bibinfo {year}
  {2021})}\BibitemShut {NoStop}%
\bibitem [{\citenamefont {Horigane}\ \emph {et~al.}(2016)\citenamefont
  {Horigane}, \citenamefont {Kihou}, \citenamefont {Fujita}, \citenamefont
  {Kajimoto}, \citenamefont {Ikeuchi}, \citenamefont {Ji}, \citenamefont
  {Akimitsu},\ and\ \citenamefont {Lee}}]{horigane2016spin}%
  \BibitemOpen
  \bibfield  {author} {\bibinfo {author} {\bibfnamefont {K.}~\bibnamefont
  {Horigane}}, \bibinfo {author} {\bibfnamefont {K.}~\bibnamefont {Kihou}},
  \bibinfo {author} {\bibfnamefont {K.}~\bibnamefont {Fujita}}, \bibinfo
  {author} {\bibfnamefont {R.}~\bibnamefont {Kajimoto}}, \bibinfo {author}
  {\bibfnamefont {K.}~\bibnamefont {Ikeuchi}}, \bibinfo {author} {\bibfnamefont
  {S.}~\bibnamefont {Ji}}, \bibinfo {author} {\bibfnamefont {J.}~\bibnamefont
  {Akimitsu}}, \ and\ \bibinfo {author} {\bibfnamefont {C.}~\bibnamefont
  {Lee}},\ }\href@noop {} {\bibfield  {journal} {\bibinfo  {journal}
  {Scientific reports}\ }\textbf {\bibinfo {volume} {6}},\ \bibinfo {pages} {1}
  (\bibinfo {year} {2016})}\BibitemShut {NoStop}%
\bibitem [{\citenamefont {Le~Tacon}\ \emph {et~al.}(2013)\citenamefont
  {Le~Tacon}, \citenamefont {Minola}, \citenamefont {Peets}, \citenamefont
  {Sala}, \citenamefont {Blanco-Canosa}, \citenamefont {Hinkov}, \citenamefont
  {Liang}, \citenamefont {Bonn}, \citenamefont {Hardy}, \citenamefont {Lin}
  \emph {et~al.}}]{le2013dispersive}%
  \BibitemOpen
  \bibfield  {author} {\bibinfo {author} {\bibfnamefont {M.}~\bibnamefont
  {Le~Tacon}}, \bibinfo {author} {\bibfnamefont {M.}~\bibnamefont {Minola}},
  \bibinfo {author} {\bibfnamefont {D.}~\bibnamefont {Peets}}, \bibinfo
  {author} {\bibfnamefont {M.~M.}\ \bibnamefont {Sala}}, \bibinfo {author}
  {\bibfnamefont {S.}~\bibnamefont {Blanco-Canosa}}, \bibinfo {author}
  {\bibfnamefont {V.}~\bibnamefont {Hinkov}}, \bibinfo {author} {\bibfnamefont
  {R.}~\bibnamefont {Liang}}, \bibinfo {author} {\bibfnamefont
  {D.}~\bibnamefont {Bonn}}, \bibinfo {author} {\bibfnamefont {W.}~\bibnamefont
  {Hardy}}, \bibinfo {author} {\bibfnamefont {C.}~\bibnamefont {Lin}},  \emph
  {et~al.},\ }\href@noop {} {\bibfield  {journal} {\bibinfo  {journal}
  {Physical Review B}\ }\textbf {\bibinfo {volume} {88}},\ \bibinfo {pages}
  {020501} (\bibinfo {year} {2013})}\BibitemShut {NoStop}%
\bibitem [{\citenamefont {Le~Tacon}\ \emph {et~al.}(2011)\citenamefont
  {Le~Tacon}, \citenamefont {Ghiringhelli}, \citenamefont {Chaloupka},
  \citenamefont {Sala}, \citenamefont {Hinkov}, \citenamefont {Haverkort},
  \citenamefont {Minola}, \citenamefont {Bakr}, \citenamefont {Zhou},
  \citenamefont {Blanco-Canosa} \emph {et~al.}}]{le2011intense}%
  \BibitemOpen
  \bibfield  {author} {\bibinfo {author} {\bibfnamefont {M.}~\bibnamefont
  {Le~Tacon}}, \bibinfo {author} {\bibfnamefont {G.}~\bibnamefont
  {Ghiringhelli}}, \bibinfo {author} {\bibfnamefont {J.}~\bibnamefont
  {Chaloupka}}, \bibinfo {author} {\bibfnamefont {M.~M.}\ \bibnamefont {Sala}},
  \bibinfo {author} {\bibfnamefont {V.}~\bibnamefont {Hinkov}}, \bibinfo
  {author} {\bibfnamefont {M.}~\bibnamefont {Haverkort}}, \bibinfo {author}
  {\bibfnamefont {M.}~\bibnamefont {Minola}}, \bibinfo {author} {\bibfnamefont
  {M.}~\bibnamefont {Bakr}}, \bibinfo {author} {\bibfnamefont {K.}~\bibnamefont
  {Zhou}}, \bibinfo {author} {\bibfnamefont {S.}~\bibnamefont {Blanco-Canosa}},
   \emph {et~al.},\ }\href@noop {} {\bibfield  {journal} {\bibinfo  {journal}
  {Nature Physics}\ }\textbf {\bibinfo {volume} {7}},\ \bibinfo {pages} {725}
  (\bibinfo {year} {2011})}\BibitemShut {NoStop}%
\bibitem [{\citenamefont {Knox}(1963)}]{knox1963theory}%
  \BibitemOpen
  \bibfield  {author} {\bibinfo {author} {\bibfnamefont {R.~S.}\ \bibnamefont
  {Knox}},\ }\href@noop {} {\bibfield  {journal} {\bibinfo  {journal} {Solid
  State Phys.}\ }\textbf {\bibinfo {volume} {5}} (\bibinfo {year}
  {1963})}\BibitemShut {NoStop}%
\bibitem [{\citenamefont {Miller}\ \emph {et~al.}(2017)\citenamefont {Miller},
  \citenamefont {Steinhoff}, \citenamefont {Pano}, \citenamefont {Klein},
  \citenamefont {Jahnke}, \citenamefont {Holleitner},\ and\ \citenamefont
  {Wurstbauer}}]{miller2017long}%
  \BibitemOpen
  \bibfield  {author} {\bibinfo {author} {\bibfnamefont {B.}~\bibnamefont
  {Miller}}, \bibinfo {author} {\bibfnamefont {A.}~\bibnamefont {Steinhoff}},
  \bibinfo {author} {\bibfnamefont {B.}~\bibnamefont {Pano}}, \bibinfo {author}
  {\bibfnamefont {J.}~\bibnamefont {Klein}}, \bibinfo {author} {\bibfnamefont
  {F.}~\bibnamefont {Jahnke}}, \bibinfo {author} {\bibfnamefont
  {A.}~\bibnamefont {Holleitner}}, \ and\ \bibinfo {author} {\bibfnamefont
  {U.}~\bibnamefont {Wurstbauer}},\ }\href@noop {} {\bibfield  {journal}
  {\bibinfo  {journal} {Nano Letters}\ }\textbf {\bibinfo {volume} {17}},\
  \bibinfo {pages} {5229} (\bibinfo {year} {2017})}\BibitemShut {NoStop}%
\bibitem [{\citenamefont {Jiang}\ \emph {et~al.}(2018)\citenamefont {Jiang},
  \citenamefont {Xu}, \citenamefont {Rasmita}, \citenamefont {Huang},
  \citenamefont {Li}, \citenamefont {Xiong},\ and\ \citenamefont
  {Gao}}]{jiang2018microsecond}%
  \BibitemOpen
  \bibfield  {author} {\bibinfo {author} {\bibfnamefont {C.}~\bibnamefont
  {Jiang}}, \bibinfo {author} {\bibfnamefont {W.}~\bibnamefont {Xu}}, \bibinfo
  {author} {\bibfnamefont {A.}~\bibnamefont {Rasmita}}, \bibinfo {author}
  {\bibfnamefont {Z.}~\bibnamefont {Huang}}, \bibinfo {author} {\bibfnamefont
  {K.}~\bibnamefont {Li}}, \bibinfo {author} {\bibfnamefont {Q.}~\bibnamefont
  {Xiong}}, \ and\ \bibinfo {author} {\bibfnamefont {W.-b.}\ \bibnamefont
  {Gao}},\ }\href@noop {} {\bibfield  {journal} {\bibinfo  {journal} {Nature
  Communications}\ }\textbf {\bibinfo {volume} {9}},\ \bibinfo {pages} {1}
  (\bibinfo {year} {2018})}\BibitemShut {NoStop}%
\bibitem [{\citenamefont {Wang}\ \emph {et~al.}(2021)\citenamefont {Wang},
  \citenamefont {Cao}, \citenamefont {Lu}, \citenamefont {Cohen}, \citenamefont
  {Kitadai}, \citenamefont {Li}, \citenamefont {Tan}, \citenamefont {Wilson},
  \citenamefont {Lui}, \citenamefont {Smirnov} \emph {et~al.}}]{wang2021spin}%
  \BibitemOpen
  \bibfield  {author} {\bibinfo {author} {\bibfnamefont {X.}~\bibnamefont
  {Wang}}, \bibinfo {author} {\bibfnamefont {J.}~\bibnamefont {Cao}}, \bibinfo
  {author} {\bibfnamefont {Z.}~\bibnamefont {Lu}}, \bibinfo {author}
  {\bibfnamefont {A.}~\bibnamefont {Cohen}}, \bibinfo {author} {\bibfnamefont
  {H.}~\bibnamefont {Kitadai}}, \bibinfo {author} {\bibfnamefont
  {T.}~\bibnamefont {Li}}, \bibinfo {author} {\bibfnamefont {Q.}~\bibnamefont
  {Tan}}, \bibinfo {author} {\bibfnamefont {M.}~\bibnamefont {Wilson}},
  \bibinfo {author} {\bibfnamefont {C.~H.}\ \bibnamefont {Lui}}, \bibinfo
  {author} {\bibfnamefont {D.}~\bibnamefont {Smirnov}},  \emph {et~al.},\
  }\href@noop {} {\bibfield  {journal} {\bibinfo  {journal} {Nature Materials}\
  }\textbf {\bibinfo {volume} {20}},\ \bibinfo {pages} {964} (\bibinfo {year}
  {2021})}\BibitemShut {NoStop}%
\end{thebibliography}%

\clearpage

\section{Supplemental Material}

\subsection{Local Projection Details}\label{a:pos}
On each site a full set of real hydrogen-like functions $s$, $p$, and $d$ were employed using the default main quantum number of the hydrogen radial function. Details of the sites on which the local projections defined in Ref.~\onlinecite{lane2020landscape} are centered within the crystal structure of NiPS$_3$ are given in Table \ref{table:NiPS3}.

\begin{table}[h]
\centering
\begin{tabular}{l|c|c|c}
\hline\hline 
NiPS$_3$      & x    & y      & z   \\\hline
S(0)& 0.2488 & 0.3345 & 1.2485\\
S(1)& 0.7512 & 0.6655 & 0.7515\\
S(2)& 0.7512 & 0.3345 & 0.7515\\
S(3)& 0.2488 & 0.6655 & 1.2485\\
S(4)& 0.7488 & 0.8345 & 1.2485\\
S(5)& 0.2512 & 0.1655 & 0.7515\\
S(6)& 0.2512 & 0.8345 & 0.7515\\
S(7)& 0.7488 & 0.1655 & 1.2485\\
S(8)& 0.25 & 0.0 & 1.247\\
S(9)& 0.75 & 0.0 & 0.753\\
S(10)& 0.75 & 0.5 & 1.247\\
S(11)& 0.25 & 0.5 & 0.753\\
P(12)& 0.5566 & 0.0 & 1.1692\\
P(13)& 0.4434 & 0.0 & 0.831\\
P(14)& 0.0566 & 0.5 & 1.1692\\
P(15)& 0.9434 & 0.5 & 0.831\\
Ni(16)& 0.0 & 0.1674 & 1.0\\
Ni(17)& 0.0 & 0.8326 & 1.0\\
Ni(18)& 0.5 & 0.6674 & 1.0\\
Ni(19)& 0.5 & 0.3326 & 1.0\\
\hline\hline
\end{tabular}
\caption{\label{table:NiPS3}
The sites on which the local projections are centered within the crystal structure of NiPS$_3$ in units of the lattice vectors.}
\end{table}

\end{document}